\documentclass[aps,prd,superscriptaddress]{revtex4}
\usepackage{epsfig,epsf}
\usepackage{amsmath}
\usepackage{amsthm}
\usepackage{amsfonts}
\usepackage{amssymb}
\usepackage{dsfont}
\usepackage{multirow}
\usepackage{appendix}
\usepackage{slashed}
\usepackage[active]{srcltx}
\usepackage{psfrag}
\usepackage{subfigure}

\begin{document}
\title{{\Large{\bf Form Factors and Differential Branching Ratio of $B \to K \mu^+ \mu^-$  in AdS/QCD  }}}

\author{\small
S. Momeni\footnote {e-mail: samira.momeni@phy.iut.ac.ir }, R. Khosravi  \footnote {e-mail: rezakhosravi @ cc.iut.ac.ir } }

\affiliation{Department of Physics, Isfahan University of
Technology, Isfahan 84156-83111, Iran }

\begin{abstract}
The holographic distribution amplitudes (DAs) for the $K$
pseudoscalar meson are derived. For this aim, the light-front wave
function (LFWF) of the $K $ meson is extracted within the framework
of the anti–de Sitter/quantum chromodynamics (AdS/QCD)
correspondence. We consider a momentum-dependent (dynamical)
helicity wave function that contains the dynamical spin effects. We
use the LFWF to predict the radius and the electromagnetic form
factor of the kaon and compare them with the experimental values.
Then, the holographic twist-2 DA of $K$ meson $\phi_{_K}(\alpha,
\mu)$ is investigated and compared with the result of the light-cone
sum rules (LCSR). The transition form factors of the semileptonic
$B\to K \ell^{+}\ell^{-}$ decays are derived from the holographic
DAs of the kaon. With the help of these form factors, the
differential branching ratio of the $B\to K\, \mu^+ \mu^-$ on $q^2$
is plotted. A comparison is made between our prediction in AdS/QCD
and the results obtained from two models including  the LCSR and the
lattice QCD as well as the experimental values.
\end{abstract}


\maketitle

\section{Introduction}
The flavor changing neutral current (FCNC) transitions have
received remarkable attention, both experimentally and theoretically.
The decay of a $b$ quark into an $s$ quark and lepton pairs, $b \to s
\ell^{+} \ell^{-}$, is a good tool to study the FCNC processes; it is
also a very good way to  probe the new physics effects beyond
the standard model (SM).

The $B \to K \ell^+ \ell^-$ decay, which occurs by the $b \to s
\ell^{+} \ell^{-}$ process at the quark level, is a suitable
candidate for experimental researchers who study the FCNC
transition. The differential branching ratio, forward-backward, and
isospin asymmetries for this transition have been measured at the
BABAR, Belle, and CDF collaborations
\cite{Wei,Aaltonen,Aaltonen1,Lees}. Researchers in the LHCb
Collaboration have reported newer results for these observable
quantities \cite{Aaij1,Aaij2,Aaij3}. Recently, the updated results
have been released for the differential branching fraction and the
angular analysis of the $B \to K \mu^+ \mu^-$ decay \cite{Bifani}.
On the other hand, physicists have tried to improve their results
for this decay via the theoretical approaches \cite{Bouchard}.
Recently, a new analysis has been made to estimate the transition
form factors of  the $B \to K \mu^+ \mu^-$ decay by the lattice QCD
\cite{Bailey}.

To evaluate the branching ratio and the other observable, we need to
describe the intended transition according to its form factors,
which are defined in terms of the distribution amplitudes (DAs). We
recall that an accurate calculation of the DAs is very important
since they provide a major source of uncertainty in theoretical
predictions. The DAs for the $K$ pseudoscalar meson have been
obtained, for the first time, from the LCSR
\cite{Chernyak,Khodjamirian1}. In recent years, a relatively new
tool named the AdS/QCD correspondence has been  used to obtain the
DAs for the light mesons. In this  approach, the wave function that
describes the hadrons in the AdS space is mapped to the wave
function used for the bound states in the light-front QCD. Both of
them satisfy a Schrodinger-like wave function equation. The
light-front DAs are derived from the holographic light-front wave
function (LFWF; for instance, see \cite{Hwang12,
Ahma1,Ahma3,Ahma4,ChaBro}).

So far, the isospin asymmetry of the $B \to K^* \mu^+ \mu^-$
transition has been considered in the AdS/QCD correspondence
\cite{Ahma5}. Dynamical spin effects have been taken into account of
the holographic pion wave function in order to predict its mean
charge radius, decay constant, the spacelike electromagnetic form
factor, twist-$2$ DA, and the photon-to-pion transition form factor
\cite{AhmaChish}. Our goal in this paper is to  extract the
twist-$2$, twist-$3$ and twist-$4$  DAs of the $K$ pseudoscalar
meson in the AdS/QCD method and use these holographic DAs to compute
the form factors and differential branching ratio for the $B \to
K\,\mu^{+}\,\mu^{-}$ transition.

Our paper is organized as follows: In Sec. II, the light-front DAs
and the holographic LFWF for the $K$  pseudoscalar meson are
calculated. In this section, the connection between the holographic
LFWF and DAs of the $K$ meson is presented. Using the holographic
DAs, the transition form factors can be investigated. In Sec. III,
we use the holographic LFWF to consider the radius and the
electromagnetic (EM) form factor of the $K$ meson and compare them
with the experimental values. We also analyze the holographic
twist-2 DA of $K$ meson $\phi_{_K}(\alpha, \mu)$  and transition
form factors of the FCNC $B\to K$ transitions. Then, the
differential branching ratio of $B\to K \mu^+ \mu^-$ decay on $q^2$
is plotted. Our prediction is compared with those made by the
lattice QCD and light-cone sum rule (LCSR) approaches, as well as
the experimental values.

\section{THE HOLOGRAPHIC DISTRIBUTION AMPLITUDES FOR THE K MESON}
The holographic DAs for the $K$ pseudoscalar meson are derived in
this section. For this aim, we plan to obtain a connection between
the DAs and the holographic LFWF of the $K$ meson.  Using the
definition of the DAs for the $K$ meson introduced by the
meson-to-vacuum matrix elements
\cite{Khodjamirian1,Chernyak,Braun,Belyaev1}, and choosing
$p^{\mu}=(p^+, \frac{m^2_{K}}{p^+}, \textbf{0}_\perp)$ for the
four-momentum of the $K$ meson, the following matrix
elements can be written in the light-front coordinate, $x^{\mu}=(x^+, x^-,
\textbf{x}_{\perp})$, at equal light-front time, $x^{+}=0$, as
\begin{eqnarray}
\langle 0 | \bar
u(0)\gamma^{\alpha}\gamma^{5}s(x^{_-})|K(p)\rangle&=&
i\,f_{_K} p^{\alpha} \int_0^1 du\,  e^{-i up^{+} x^{-}} \phi_{_K}(u, \mu)\,,\label{eq21}\\
\langle 0 | \bar u(0) \gamma^{5} s(x^{_-})|K(p)\rangle&=&
-i\,\frac{f_{_K} m^{2}_{_K}}{m_u+m_s} \int_0^1 du\, e^{-i up^{+} x^{-}}  \phi_{\rho}(u, \mu)\,,\label{eq22}\\
\langle 0 | \bar u(0) \sigma^{\alpha \beta} (1+\gamma^{5})
s(x^{_-})|K(p)\rangle&=& \frac{i}{6}\,\frac{f_{_K}
m^{2}_{_K}}{(m_u+m_s)}\,p^{[{\alpha}}x^{\beta]} \int_0^1 du\, e^{-iu
p^{+} x^{-}}
\phi_{\sigma}(u, \mu)\,,\label{eq23}\\
\langle 0 | \bar u(0)\gamma^{\alpha} s(x^{_-})|K(p)\rangle&=&
i\,f_{_K} {(x^-)}^{2} p^{\alpha} \int_{0}^{1} du\, e^{-i up^{+}
x^{-}} g_{1}(u, \mu) -f_{_K}
(x^{\alpha}-\frac{x^-}{p^{+}} p^{\alpha})\nonumber\\
&\times& \int_{0}^{1} du\, e^{-i up^{+} x^{-}} g_{2}(u, \mu)\,,
\label{eq24}
\end{eqnarray}
where $\mu$ is the renormalization scale and  $f_{_K}$ is the decay
constant of the $K$ pseudoscalar meson. In these relations,
$\phi_{_K}$ is twist-2, $\phi_{\rho}$ and $\phi_{\sigma}$  are
twist-3, and $g_{1}$ and $g_{2}$ are twist-4 DAs for the $K$ meson.
To isolate  $\phi_{_K}$ and $\phi_{\rho}$, we take $\alpha=+$ and
apply the Fourier transform of  Eqs. (\ref{eq21}) and (\ref{eq22})
with respect to $x^{-}$. It yields
\begin{eqnarray}
\phi_{_K}(\alpha, \mu)&=&-\frac{i}{f_{K}}\int dx^{-}\,e^{i\alpha
p^{+} x^{-}} \langle 0 | \bar
u(0)\gamma^{\alpha}\gamma^{5}s(x^{_-})|K(p)\rangle
,\label{eq25}\\
\phi_{\rho}(\alpha, \mu)&=&i\frac{(m_u+m_s)}{f_{_K} m^{2}_{_K}}p^{+}
\int dx^{-}\,e^{i\alpha p^{+} x^{-}} \langle 0 | \bar u(0)
\gamma^{5} s(x^{_-})|K(p)\rangle.\label{eq26}
\end{eqnarray}
Choosing $\sigma^{+-}$ in Eq. (\ref{eq23}), and using integration by
parts with the boundary condition $\phi(u)|^{1}_{0}=0$, as well as
performing the Fourier transform with respect to $x^{_-}$, the derivative of the twist-3 $\phi_{\sigma}(\alpha, \mu)$ is obtained
as
\begin{eqnarray}\label{eq27}
\frac{\partial \phi_{\sigma}(\alpha, \mu)}{\partial
\alpha}&=&\frac{6 (m_u+m_s)}{f_{_K} m^2_{_K}} p^{+} \int dx^{-} e^{i
\alpha p^{+} x^{-}} \langle 0 | \bar u(0) \sigma^{+ -} (1+
\gamma^{5}) s(x^{-})|K(p)\rangle.
\end{eqnarray}
Taking $\alpha=+$ (and afterwards $\alpha=-$)  in Eq. (\ref{eq24}),
and then using integration by part, the following relations are
derived:
\begin{eqnarray}
\langle 0 | \bar u(0)\gamma^{+} s(x^{-})|K(p)\rangle &=& \frac{i
f_{_K}}{p^{+}}\Bigg[\int_{0}^{1} du\, e^{-i up^{+} x^{-}}\,
\frac{\partial^2 g_{1}(u, \mu)}{\partial u^2} -\int_{0}^{1} du\,
e^{-i up^{+} x^{-}}\,\frac{\partial g_{2}(u, \mu)}{\partial
u}\Bigg], \label{eq28}\\
\langle 0 | \bar u(0)\gamma^{-} s(x^{-})|K(p)\rangle &=& \frac{i
f_{_K}}{p^{+}}\Bigg[\frac{m_K^2}{(p^+)^2} \int_{0}^{1} du\, e^{-i
up^{+} x^{-}}\, \frac{\partial^2 g_{1}(u, \mu)}{\partial u^2}\,
-\left(1-\frac{m_{K}^2}{(p^{+})^2}\right)\nonumber\\
&\times& \int_{0}^{1} du\, e^{-i up^{+} x^{-}}\,\frac{\partial
g_{2}(u, \mu)}{\partial u}\Bigg].\,\label{eq29}
\end{eqnarray}
Solving Eqs. (\ref{eq28}) and (\ref{eq29}) in terms of
$\frac{\partial^2 g_{1}(u, \mu)}{\partial u^2}$ and $\frac{\partial
g_{2}(u, \mu)}{\partial u}$, as well as  performing the Fourier
transform with respect to $x^{-}$, we obtain
\begin{eqnarray}
\frac{\partial g_{2}(\alpha, \mu)}{\partial
\alpha}&=&\frac{i}{f_{K}[2
m_{K}^2-(p^{+})^2]}\int dx^{-} e^{i \alpha p^{+} x^{-}}\nonumber\\
&\times&\left[\frac{m_{K}^2}{(p^{+})^2}  \langle 0 | \bar
u(0)\gamma^{+} s(x^{-})|K(p)\rangle - \langle 0 | \bar
u(0)\gamma^{-} s(x^{-})|K(p)\rangle\right],\label{eq210}\\
\frac{\partial^2 g_{1}(\alpha, \mu)}{\partial
\alpha^2}&=&\frac{i}{f_{K}[2 m_{K}^2-(p^{+})^2]}\int dx^{-} e^{i
\alpha p^{+}
x^{-}}\nonumber\\&\times&\left[(\frac{m_{K}^2}{(p^{+})^2}-1) \langle
0 | \bar u(0)\gamma^{+} s(x^{-})|K(p)\rangle - \langle 0 | \bar
u(0)\gamma^{-} s(x^{-})|K(p)\rangle\right].\label{eq211}
\end{eqnarray}

In order to evaluate the holographic DAs for the $K$ meson, the
hadronic matrix elements should be determined  in Eqs. (\ref{eq25})
-(\ref{eq27}) and (\ref{eq210})-(\ref{eq211}). For this purpose, the
Fock expansion of noninteracting two-particle states is used for a
hadronic eigenstate $|P\rangle$ as \cite{R1}
\begin{eqnarray}\label{eq212}
|P(p)\rangle=\sqrt{4\pi N_c} \sum_{h,\bar{h}} \int \frac{d k^{+}
d^2{\bf {k}_{\bot}}}{16\pi^3 \sqrt{k^{+}(p^{+}-k^{+})}}
\Psi^{P}_{h,\bar{h}} ( \frac{k^+}{p^+}, {\bf {k}_{\bot}} )
|k^+,\textbf{k}_{\bot},h;\,p^+-k^+,-\textbf{k}_{\bot},\bar{h}\rangle,
\end{eqnarray}
in which  $\Psi^{P}_{h,\bar{h}}( \alpha, {\bf {k}_{\bot}} )$  is the
LFWF of the pseudoscalar meson, and $h$ and $\bar{h}$ are the
helicities of the quark and anti-quark, respectively. By utilizing
the expansion of Dirac fields (quark and antiquark) in terms of
particle creation and annihilation operators, and also the equal
light-front time anticommination relations for these operators, the
matrix element $\langle 0 | \bar{u}(0)\, \Gamma\, s(x^-) |P(p)
\rangle$ is obtained as
\begin{eqnarray}\label{eq213}
\langle 0 | \bar{u}(0)\, \Gamma\, s(x^-) |P(p) \rangle&=& \sqrt{4\pi
N_c} \sum_{h,\bar{h}} \int \frac{d k^{+}d^2{\bf {k}_{\bot}}
\,\Theta(|\mathbf{k}_{\bot}| <\mu)}
{16\,\pi^3\sqrt{k^+(p^{+}-k^{+})}}
\Psi^{P}_{h,\bar{h}}( \alpha, {\bf {k}_{\bot}} )\nonumber \\
&& \times \bar{v}_{\bar{h}}(p^{+}-k^{+},-{\bf {k}_{\bot}})\,
\Gamma\, u_h(k^+,{\bf {k}_{\bot}}) e^{-ik^{+}x^{-}}\,,
\end{eqnarray}
in which  $ u_h$ and $v_{h}$ are light-front helicity spinors for
the quark and antiquark, respectively. The renormalization scale
$\mu$ is used as the ultraviolet cutoff on transverse momenta
\cite{Kogut, Diehl}. In our work, $\Gamma$ can be
$\sigma^{+-}(1+\gamma^{5})$, $\gamma^{+}$, or $\gamma^{-}$. By
integrating with respect to $k^{+}$ and applying the Fourier
transform to the left and right- hand sides of Eq. (\ref{eq213}),
the following result is obtained:
\begin{eqnarray}\label{eq214}
\int d x^- e^{i\alpha p^+x^-} \langle 0 | \bar{u}(0)\,  \Gamma\,
s(x^-)|P(p) \rangle &=& \frac{\sqrt{4\pi N_c}}{p^+}\sum_{h,\bar{h}}
\int^{|\mathbf{k}_{\bot}| < \mu}
\frac{d^2\mathbf{k}_{\bot}}{(2\pi)^3}\Psi^{P}_{h,\bar{h}}(\alpha,\mathbf{k}_{\bot})\\
\nonumber &&\times \left \{ \frac{\bar{v}_{\bar{h}}(\bar
\alpha\,p^{+},-\mathbf{k}_{\bot})}{\sqrt{\bar\alpha}} \,\Gamma\,
\frac{u_h(\alpha\, p^+,\mathbf{k}_{\bot})}{\sqrt{\alpha}} \right
\}\,,
\end{eqnarray}
where $\alpha=\frac{k^+}{p^+}$, and $\bar \alpha=1-\alpha$. In the
$\mathbf{k}$ space, the holographic LFWF is given  as \cite{R1}
\begin{eqnarray}\label{eq2155}
\Psi^{P}_{h,\bar{h}}(\alpha,\mathbf{k}_{\bot})=\frac{1}{\sqrt{4\pi}}
S^{P}_{h,\bar{h}}(\alpha,\mathbf{k}_{\bot})
\phi(\alpha,\mathbf{k}_{\bot}).
\end{eqnarray}

The structure of $S^{P}_{h,\bar{h}}(\alpha,\mathbf{k}_{\bot})$ for
the pseudoscalar mesons that includes the helicity-dependent wave
function is as follows:
\begin{eqnarray}\label{eq215}
S_{h,\bar{h}}^{P}(\alpha,\mathbf{k}_{\bot})= \frac{\bar{u}_{h}
(\alpha\,p^+ ,-\mathbf{k}_{\bot} )}{\sqrt{\alpha}} \,\left[(A\not\!p
+ B m_{_K})\gamma^{5}\right] \frac{v_{\bar{h}}(\bar\alpha\,
p^+,\mathbf{k}_{\bot})}{\sqrt{\bar\alpha}}\,,
\end{eqnarray}
where $A$ and $B$ are arbitrary constants. If $B \neq 0$, the
dynamical spin effects are allowed. For considering the dynamical
spin effects, $A$ and $B$ are usually taken in two cases: $(A = 0;B
= 1)$ and $(A = 1;B = 1)$
\cite{AhmaChish,Heinzl,Choi,Trawiski,ChaBro}.

Using the light-front spinors presented in Ref. \cite{Brodsky6},
$S_{h,\bar{h}}^{P}$ is evaluated for the $K$ meson as
\begin{eqnarray}\label{eq216}
i\,S_{h,\bar{h}}^{K}(\alpha,\mathbf{k}_{\bot}) &=& \mp \frac{
A}{\alpha \bar\alpha}\Bigg\{\left[\alpha \bar\alpha\, m_{_K}^{2}+
m_{u}m_{s}+ {k}^{2} \right] \delta_{h\pm,\bar{h}\mp} \pm {k}
\left[m_{u}e^{ -i\theta_k}\delta_{h+,\bar{h}+}+m_{s}e^{
i\theta_k}\delta_{h-,\bar{h}-}\right] \Bigg\}\nonumber\\
&\mp& \frac{B\, m_{_K}}{\alpha \bar\alpha} \left[\alpha\,m_{s}+ \bar
\alpha\, m_{u}\mp {k} e^{\mp i \theta_k}\right]
\delta_{h\pm,\bar{h}\mp}\,,
\end{eqnarray}
where  $k e^{\pm i \theta_k}$ is the complex form of the transverse
momentum $\mathbf{k}_{\bot}$; in addition, $h +$ and $h-$ are used
for positive and negative helicity, respectively.

The light-front spinors are also utilized to obtain the matrix
elements in the right-hand side of Eq. (\ref{eq214}). The final
results can  be written as
\begin{eqnarray}
\frac{\bar{v}_{\bar{h}}}{\sqrt{\bar{\alpha}}}\gamma^+\frac{u_h}{\sqrt{\alpha}} &=&2 p^+ \delta_{h\pm,\bar{h}\mp}\,,\nonumber\\
\frac{\bar{v}_{\bar{h}}}{\sqrt{\bar{\alpha}}}\gamma^-\frac{u_h}{\sqrt{\alpha}}
&=&2 p^+ \delta_{h\pm,\bar{h}\mp}\,,\nonumber\\
\frac{\bar{v}_{\bar{h}}}{\sqrt{\bar{\alpha}}} \gamma^{+}\gamma^{5} \frac{u_h}{\sqrt{\alpha}}&=&\pm\, 2p^+ \delta_{h\pm,\bar{h}\mp}\,,\nonumber\\
\frac{\bar{v}_{\bar{h}}}{\sqrt{\bar{\alpha}}} \gamma^{5}
\frac{u_h}{\sqrt{\alpha}}&=& \frac{1}{\alpha\,\bar{\alpha}}\left\{ k
e^{\pm i \theta_k}\delta_{h\pm,\bar{h}\pm}
\mp\left(\alpha\, m_{s}+\bar{\alpha}\,m_{u}\right) \delta_{h\pm,\bar{h}\mp} \right\},\nonumber\\
\frac{\bar{v}_{\bar{h}}}{\sqrt{\bar{\alpha}}}
\sigma^{+-}(1+\gamma^{5}) \frac{u_h}{\sqrt{\alpha}}
&=&\frac{4\,i}{\alpha\,\bar{\alpha}} \left\{\mp k e^{\pm i
\theta_k}\,(1-2\,\alpha) \delta_{h\pm,\bar{h}\pm}
+\alpha\,m_{{u}}\,\delta_{h+,\bar{h}-} + \bar{\alpha}\,m_{s}
\,\delta_{h-,\bar{h}+}\right\}. \label{eq217}
\end{eqnarray}

Inserting  Eqs. (\ref{eq216})-(\ref{eq217}) in Eq. (\ref{eq214}),
the hadronic matrix elements in Eqs. (\ref{eq25})-(\ref{eq27}) and
(\ref{eq210})-(\ref{eq211}) are determined. Therefore, the
holographic DAs can be calculated for the $K$ meson in terms of
$\phi(\alpha, \mathbf{k}_{\bot})$ in the $\textbf{k}$ space.
Applying the Fourier transform to $r$ space and using relations such
as $\int_{0}^{2\pi} e^{-i  k r cos\theta} d\theta =2\pi J_{0}(k r)$,
and $\int_{0}^{\mu} k\, J_{0}(k r)\,d(k r) ={\mu}/{r}\,J_{1}(\mu
r)$, where $J_{0}$ and $J_{1}$ are Bessel functions, the following
expressions are obtained for the holographic DAs in the $r$ space:
\begin{eqnarray}\label{eq218}
\phi_{_K}(\alpha, \mu) &=& \frac{\beta_{1}}{\alpha \bar{\alpha}}
\int {dr}\,\mu J_{1}(\mu r) \left\{2 A \left(\alpha \bar{\alpha}
m_{_K}^{2} + m_{u} m_{s} - \nabla^{2}\right )+ B m_{_K}
\left(\bar{\alpha}\,m_u + \alpha\, m_{s}\right)\right\} \phi(\alpha,
r),
\nonumber\\ \nonumber\\
\phi_{\rho}(\alpha, \mu) &=& -\frac{(m_{s}+m_{u}) \beta_{1}}{
\alpha^2 \bar{\alpha}^2\, m_{_K}^2} \int {dr}\,\mu J_{1}(\mu r)\Big
\{A\left[(\alpha\,m_{u}+\bar{\alpha}\,m_{s}) \left(\alpha
\bar{\alpha} m_{_K}^{2}+ m_{u}
m_{s}-\nabla^2\right)\right.\nonumber\\&-&\left.(m_{u}+m_{s})\nabla^2
\right]- B m_{_K}\left[\left(\alpha
m_{s}+\bar{\alpha}m_u\right)^2-\nabla^2\right] \Big\} \,\phi(\alpha,
r),
\nonumber\\ \nonumber\\
\frac{\partial \phi_{\sigma}(\alpha, \mu)}{\partial
\alpha}&=&-\frac{24 (m_u+m_s) \beta_{1}}{\alpha^2 \bar{\alpha}^2\,
m_{_K}^{2}} \int {dr}\,\mu J_{1}(\mu r)\Big
\{A\left[(\alpha\,m_{u}-\bar{\alpha}\,m_{s}) \left(\alpha
\bar{\alpha} m_{_K}^{2}+ m_{u}
m_{s}-\nabla^2\right)\right.\nonumber\\&-&\left.(2\alpha-1)(m_{u}-m_{s})\nabla^2
\right]+ B m_{_K}\left[ \alpha^2 m^2_{u}-\bar{\alpha}^2 m^2_s -(2
\alpha-1) \nabla^2\right] \Big\} \,\phi(\alpha, r),
\nonumber\\ \nonumber\\
\frac{\partial g_{2}(\alpha, \mu)}{\partial \alpha}&=&\frac{\beta_1
\beta_{2}} {\beta_{3}\,\alpha\bar{\alpha} m_{_K} } \int {dr}\,\mu
J_{1}(\mu r)\,B \left[ \alpha\, m_{\bar{u}}+\bar\alpha\, m_{s}
\right] \phi(\alpha, r),
\nonumber\\ \nonumber\\
\frac{\partial^2 g_{1}(\alpha, \mu)}{\partial \alpha^2}&=&
\frac{\beta_1 (\beta_{2}-1)} {\beta_{3}\,\alpha\bar{\alpha} m_{_K} }
\int {dr}\,\mu J_{1}(\mu r)\, B\left[ \alpha\, m_{u}+\bar\alpha\,
m_{s}) \right] \phi(\alpha, r),
\end{eqnarray}
where $\sqrt{N_{c}}/(\pi\,f_{K})=\beta_{1}$,
$\left[1-{m_{_K}^2}/{{(p^+)}^{2}}\right]=\beta_{2}$ and
$\left[2-{{(p^+)}^{2}}/{m_{_K}^2}\right]=\beta_{3}$.

To specify $\phi (\alpha, r)$, which includes dynamical
properties of $K$ in the LFWF, we are going to use the AdS/QCD.
Based on a first semiclassical approximation to the light-front QCD,
with massless quarks,  function $\phi$ can be factorized as
\cite{GFdeSJBr}
\begin{eqnarray}\label{eq219}
\phi(\zeta,\alpha, \theta) =\mathcal{N} \frac{\psi(\zeta)}{\sqrt{2
\pi \zeta}}\, f(\alpha) \, e^{i L \theta},
\end{eqnarray}
where $\mathcal{N}$ is a normalization constant. In this relation,
$L$ is the orbital angular momentum quantum number and variable
$\zeta=\sqrt{\alpha(1-\alpha)}\,r$, where $r$ is the transverse
distance between the quark and antiquark forming the meson. Function
$\psi(\zeta)$ satisfies the so-called holographic light-front
Schrodinger-like equation as
\begin{eqnarray}\label{eq220}
\left(-\frac{d^2}{d\zeta^2}-\frac{1-4L^2}{4\zeta^2}+U(\zeta)\right)\psi(\zeta)=M^2
\psi(\zeta),
\end{eqnarray}
where $M$ is the hadron bound-state mass and $U(\zeta)$ is the effective
potential. It should be noted that all the interaction terms and the
effects of higher Fock states on the valence ($N = 2$ for mesons)
state are hidden in the confinement potential.

According to  the AdS/QCD, the holographic light-front Schrodinger
equation is mapped onto the wave equation for strings propagating in
the AdS space if $\zeta$ is identified with the fifth dimension in
AdS space. To illustrate this issue, the invariant action (up to
bilinear terms) is written for a scalar field in the AdS$_{5}$ space
as
\begin{eqnarray} \label{eq221}
S = \frac{1}{2} \int \! d^4 x \, dz  \,\sqrt{g} \,e^{\varphi(z)}
\left( g^{MN} \partial_M \Phi \partial_N \Phi -  \mu^2 \Phi^2
\right) ,
\end{eqnarray}
where $g = {(\frac{R}{z})}^{10}$ is the modulus of the determinant
of the metric tensor $g_{MN}$. Moreover, $\Phi (x^\mu, z)$ is a
scalar field. Mass $\mu$ in Eq. (\ref{eq221}) is not a physical
observable. In this action, the dilaton background $\varphi(z)$ is
only a function of the holographic variable $z$ that vanishes if $z
\to \infty$. Variation of Eq. (\ref{eq221})  and making the ansatz
$\Phi(x^\mu, z) = e^{-i P \cdot x} \Theta(z)$, which describe a free
hadronic state with four-momentum $P$ in holographic QCD, the
eigenvalue equation is obtained as
\begin{eqnarray} \label{eq222}
\left[-\frac{ z^{3}}{e^{\varphi(z)}}   \partial_z
\left(\frac{e^\varphi(z)}{z^{3}} \partial_z\right) + \left(\frac{\mu
R}{z}\right)^2\right] \Theta(z) = {M}^2 \Theta(z),
\end{eqnarray}
where  $P_\mu P^\mu = M^2$ is the invariant mass. Factoring out the
scale $(1/z)^{-\frac{3}{2}}$ and dilaton factors from the AdS field
as $\Theta=({\frac{R}{z}})^{-\frac{3}{2}}e^{-\varphi(z)/2}\psi(z)$,
and using a substitution as $z \to \zeta$, the  light-front
Schrodinger equation [Eq. (\ref{eq220})] is fined with the effective
potential $ U(\zeta) = \frac{1}{2} \varphi''(\zeta) +\frac{1}{4}
\varphi'(\zeta)^2  - \frac{1}{\zeta} \varphi'(\zeta)$,  and the AdS
mass  $(\mu R)^2 = L^2-1$. In this correspondence, $ \varphi(\zeta)$
and $(\mu R)^2$ are related to the effective potential and the
internal orbital angular momentum $L$, respectively.

Choosing $\varphi(\zeta)=\kappa^2 \zeta^2$ in the soft-wall model
\cite{Karch} leads to $ U(\zeta)=\kappa^4 \zeta^2-2\kappa^2$. It
should be noted that the harmonic form of this potential is unique
that is the most remarkable feature of the light-front holographic
QCD \cite{Teramond}. Solving Eq. (\ref{eq222}) with this potential
and comparing the equation with the quantum mechanical oscillator in
the polar coordinates, the results are obtained for eigenfunctions
[$\psi_{n, L}(\zeta)$] and eigenvalues [$ M^2(n, L, S)$].

Therefore, $\phi (r, \alpha)$ for the $K$ meson with massless
quarks, and $n=0$,  $L = 0 $, is obtained as
\begin{eqnarray}\label{eq223}
\phi(\alpha,\zeta)= \mathcal{N}
\frac{\kappa}{\sqrt{\pi}}\sqrt{\alpha \bar\alpha} \exp
\left(-\frac{\kappa^2 \zeta^2}{2}\right),
\end{eqnarray}
where $\kappa$ is the AdS/QCD scale. It should be noted that the
condition $ \int_{0}^{1} d\alpha
\frac{f(\alpha)^2}{\alpha\,\bar\alpha}=1 $ is used to determine the
function $f(\alpha)$ in Eq. (\ref{eq219}) \cite{GFdeSJBr}. To
include the mass of quarks in Eq. (\ref{eq223}), first, a Fourier
transform is applied to $\textbf{k}$ space as
$\widetilde{\phi}(\alpha, \textbf{k}_{\bot})=\int\,
d^2\textbf{r}\,e^{-ikr\cos\theta_k}\,\phi(\alpha,\zeta)$; it yields
\begin{eqnarray}\label{eq224}
{\widetilde{\phi}}(\alpha, \textbf{k}_{\bot})= \mathcal{N}\,
\frac{2}{ \sqrt{\alpha\bar\alpha\,}}\,\frac{\sqrt{\pi}}{\kappa}
\exp\left(-\frac{k^2}{2\alpha\bar\alpha\,\kappa^2}\right);
\end{eqnarray}
then, this substitution is used \cite{BroTera5},
\begin{eqnarray}\label{eq225}
\frac{k^2}{\alpha\bar\alpha}\to \frac{k^2}{\alpha\bar\alpha}+
\frac{m_{u}^2}{\alpha}+\frac{m^2_{s}}{\bar\alpha}.
\end{eqnarray}
After substituting this into the wave function and Fourier
transforming back to the transverse position space, the final form
of the AdS/QCD wave function is obtained as
\begin{eqnarray}\label{eq226}
\phi (\zeta, \alpha) =\mathcal{N}\, \frac{\kappa}{\sqrt{\pi}} \,
\sqrt{\alpha\,\bar\alpha} \exp \left(-\frac{\kappa^2
\zeta^2}{2}\right) \exp\left
\{-\left[\frac{\bar\alpha\,m_u^2-\alpha\,m^2_{s})}{2\alpha
\bar\alpha\,\kappa^2} \right] \right \}.
\end{eqnarray}
In position space, $\mathcal{N}$ can be fixed by this normalization
condition \cite{R1},
\begin{equation}\label{eq227}
\int d^{2} {\mathbf{r}} \,d\alpha \Bigg[\sum_{h,\bar{h}} |\Psi^{K
}_{h,\bar{h}}(r, \alpha)|^{2}\Bigg] = 1.
\end{equation}

\section{Numerical analysis}
In this section, we present our numerical analysis for the
light-front holographic DAs of  the $K$ meson,  the $B \to K
\ell^{+}\ell^{-}$ transition form factors, as well as the
differential branching ratio of the $B \to K \mu^{+}\mu^{-}$ transition
on $q^2$.

According to the light-front holographic prediction, the mass
squared of mesons composed of light quarks is given as $ M^2(n, L,
S)=4\,\kappa^2\,(n+L+\frac{S}{2})$, where the quantum numbers $L$
and $n$ describe the orbital angular momentum and excitations of the
meson spectrum, respectively. By fitting this mass relation to the
experimentally measured Regge slopes, the AdS/QCD scale $\kappa$ is
reported to be $590~\rm{MeV}$ for pseudoscalar mesons
\cite{Teramond}. In this paper, we choose $\kappa=590~\rm{MeV}$ in
our analysis. In addition, we consider  two sets for $A$ and $B$ as
set I  $(A = 1;B = 1)$ and set II  $(A = 0;B = 1)$ that allow for
considering the dynamical spin effects.

\begin{figure}[th]
\includegraphics[width=6 cm,height=6 cm]{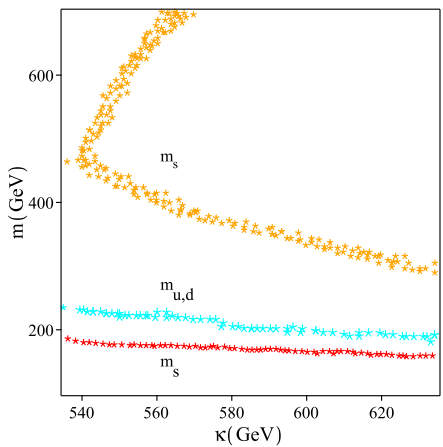}
\caption{The available spaces for the quark masses $m_{u, d, s}$
under the constraints from the experimental values of the decay
constants $f_{\pi}$ and $f_{K}$.}\label{F1}
\end{figure}

Using the experimental values of the decay constants, $f_{\pi}$ and
$f_{K}$, and choosing  the value of $\kappa$, we can obtain the mass
of the light quarks related to our analysis; they are in fact  the
effective quark masses used in the holographic LFWFs
\cite{Teramond}. The decay constant for a pseudoscalar meson, which
contains  $q$ and $q'$ quarks, can be defined as
\begin{equation}\label{eq31}
\langle
0|\bar{q}(0)\gamma^{\alpha}\,\gamma^{5}\,q'(0)|S(p)\rangle=if_{S}\,p^{\alpha}.
\end{equation}
After expanding the left-hand side of Eq. (\ref{eq31}) in the
procedure described in the previous section, the decay constant
formula for the pion and kaon in the AdS/QCD correspondence is
calculated as
\begin{eqnarray}\label{eq32}
f_{S}&=&\frac{\sqrt{N_{c}}}{\pi}\int_{0}^{1}d\alpha\, \left[ B
\left(\bar{\alpha}\,m_{\bar q} + \alpha\, m_{q'}\right) m_{S}+ 2 A
\left(\alpha \bar{\alpha}~ m_{S}^{2} + m_{\bar q} m_{q'} -
\nabla^{2}\right ) \right] \frac{\phi (\alpha,r)}{{\alpha
\bar{\alpha}}}\Bigg|_{r=0}.
\end{eqnarray}
The effective masses for two light quarks, $u$ and $d$, are equal in
the AdS/QCD. So, by inserting $m_u=m_d$, in addition to the
experimental value $f_{\pi}=130 \pm 0.26~\rm{MeV}$, and $(A=1;B=1)$
in Eq. (\ref{eq32}), we can plot $m_u$ with respect to $\kappa$ in
the region between $535< \kappa <635$ (see Fig. \ref{F1}). By having
the values of $m_{u}$ according to $\kappa$, as well as the
experimental value $f_{K}=156 \pm 0.49~\rm{MeV}$, and applying them
in Eq. (\ref{eq32}), we can also display $m_s$ based on $\kappa$,
numerically. It is obvious that the $s$ quark mass must be larger
than the mass of $u$ and $d$ quark. In addition, we consider
$700~\rm{GeV}$ as an upper limit for $m_s$.  Our numerical analysis
shows that for each value of $\kappa$ between
$537\leq\kappa\leq567$, there are three values for $m_s$, one
unacceptable (red star) and two acceptable (orange  star). For each
value of $\kappa>567$, there is only one acceptable value that is
smaller than the upper limit. According to Fig. \ref{F1}, for
$\kappa=590~\rm{MeV}$, the mass of quarks $[m_{u,d}, m_s]$ is
obtained in $\rm{MeV}$ as $[200, 350]$.

We choose $(A=0;B=1)$,  repeat the previous steps, and obtain that
the mass of quarks $[m_{u,d}, m_s]$ is $[100, 220]$ in $\rm{MeV}$.

Using the holographic LFWF,  the kaon radius observable is predicted
for two sets $(A=1;B=1)$ and $(A=0;B=1)$. This observable is
sensitive to long-distance (LD; nonperturbative) physics. The
root-mean-square kaon radius is given by \cite{BroTera6}
\begin{equation}\label{eq303}
r_{K} = \left[\frac{3}{2} \int d^{2} {\mathbf{r}} \,d\alpha~  (r
\bar \alpha )^2 |\Psi^{K}(r,\alpha)|^2 \right]^{1/2}\,,
\end{equation}
where
\begin{eqnarray}
|\Psi^{K}(r, \alpha)|^2=\sum_{h, \bar{h}}|\Psi^{K}_{h, \bar{h}}(r,
\alpha)|^2.
\end{eqnarray}
Our predictions  for $r_{K}$ are presented in Table \ref{T301}. As
can be seen, we get a better agreement with the experimental value
for the spin-improved LFWF using set I. Our prediction for set II is
closer to that via the lattice QCD.
\begin{table}[th]
\caption{Predictions for $K$ meson radius via the lattice QCD and
AdS/QCD correspondence.}\label{T301}
\begin{ruledtabular}
\begin{tabular}{ccccc}
& Ours $(A=1;B=1)$& Ours $(A=0;B=1)$  & Lattice QCD \cite{Aoki}& Exp \cite{pdgk}    \nonumber     \\
\hline
Value (fm)& $0.52 \pm 0.07$ & $0.63 \pm 0.09$  &   0.62 &$0.56\pm0.03$ \nonumber     \\
\end{tabular}
\end{ruledtabular}
\end{table}

\begin{figure}
\includegraphics[width=7.5cm,height=7cm]{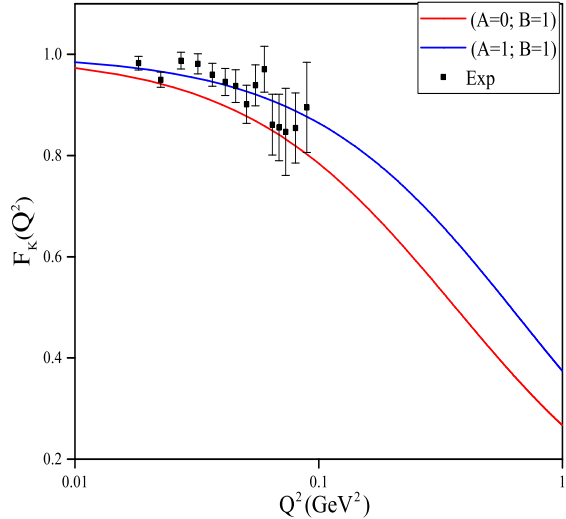}
\caption{Our predictions and  experimental data for the  EM  form
factor of the $K$ meson.} \label{F302}
\end{figure}

For a better analysis of the holographic LFWF, we investigate the
behavior of  the EM form factor for the $K$ meson in the AdS/QCD
approach. The kaon EM form factor is defined as
\begin{eqnarray}
\langle K(p)|J^{EM}_{\mu}(0)|K(p')\rangle=2\,(p+p')_{\mu}F_{K}(Q^2),
\end{eqnarray}
where $(p-p')^2=-Q^2$. The EM current is $J^{EM}_{\mu}=\frac{2}{3}\,
\bar{u}(0)\gamma_{\mu}\,u(0)-\frac{1}{3}\,
\bar{s}(0)\gamma_{\mu}\,s(0)$. The EM form factor  can be expressed
in terms of the LFWF as \cite{Drell,West}
\begin{eqnarray}
F_{K}(Q^2)=\int\,d^{2} {\mathbf{r}} \,d\alpha\,J_{0}[(r \bar \alpha)
Q]|\Psi^{K}(r, \alpha)|^2.
\end{eqnarray}
Our predictions and the experimental data \cite{Amendolia} for the
EM form factor of the $K$ meson with respect to $Q^2$, in the
interval $0.10 \,\rm {GeV}^2 \leq  Q^2 \leq 1\, \rm {GeV}^2$, are
shown in Fig. (\ref{F302}).
As can be seen, our predictions for two sets are in a satisfactory
agreement with the experimental data.

\begin{figure}[!ht]
\includegraphics[width=6 cm,height=6 cm]{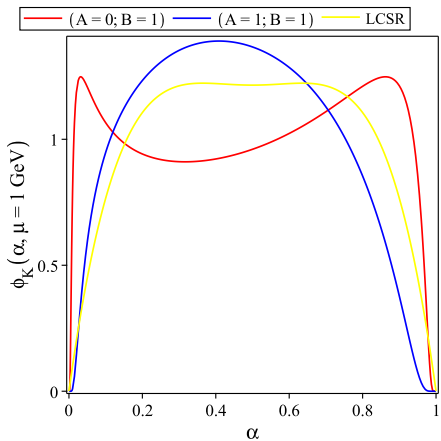}
\caption{The results for $\phi_{K}$ at $\mu=1~\rm{GeV}$ with the
AdS/QCD and LCSR.  }\label{F2}
\end{figure}

Figure \ref{F2} shows the holographic twist-2 DA $\phi_{_K}(\alpha,
\mu)$ with respect to $\alpha$, obtained form Eqs. (\ref{eq218}), on
which red and blue lines show the results for two sets in
$\mu=1~\rm{GeV}$, respectively. In this figure, we compare the
holographic twist-2 DA  with the prediction of the LCSR. It can be
seen that  $\phi_{_K}(\alpha)$ for set II is broader than  both
predictions  for set I and the LCSR.

The moments $\langle \xi_{n} \rangle$ and  inverse moment
$\left\langle {\alpha}^{-1} \right\rangle$  can be investigated
based on the twist-2 DA $\phi_{_K}(\alpha,\mu)$ as
\begin{eqnarray}
\langle \xi_{n} \rangle &=& \int_0^1 d \alpha (2\alpha-1)^n
\phi_{_K}(\alpha,\mu),\nonumber\\
\left\langle {\alpha}^{-1} \right\rangle &=& \int_0^1 d
\alpha\frac{\phi_{_K}(\alpha,\mu)}{\alpha}\,.
\end{eqnarray}
By using the holographic DA $\phi_{_K}(\alpha,\mu)$, we calculate
$\langle \xi_{2} \rangle$, $\langle \xi_{4} \rangle$, and $\langle
{\alpha}^{-1} \rangle$ and compare them with the predictions of some
nonperturbative methods such as the light-front quark model (LFQM),
lattice QCD, and LCSR. Our results are presented in Table
\ref{T302}.
\begin{table}
\caption{Prediction values for $\langle \xi_{2} \rangle$, $\langle
\xi_{4} \rangle$, and $\langle {\alpha}^{-1} \rangle$ via some
methods.}\label{T302}
\begin{ruledtabular}
\begin{tabular}{ccccc}
DA   & $\mu$ [GeV] &$\langle \xi_2 \rangle$ &$\langle \xi_4 \rangle$&$\langle \alpha^{-1} \rangle$ \\
\hline
Ours $(A=1;B=1)$ &$ 1$  & $0.21\pm 0.02$ &$0.09\pm 0.01$&$3.54\pm 0.42$\\
Ours $(A=0;B=1)$ &$ 1$  & $0.32\pm 0.04$ &$0.18\pm 0.02$&$5.33\pm 0.74$\\
LFQM \cite{Choi} & $ 1$ & $0.21$ &$0.09$ &-- \\
LFQM \cite{Chung92}& $1$ & $0.20$ & $0.08$&--\\
Lattice \cite{nam17} & $ 1$ & $0.20$ &0.09 &--\\
Lattice \cite{Braun6} & $2$ & $0.26$ &-- &--\\
LCSR \cite{Ball5}  & $2$ & $0.26\pm 0.04$ &-- &--\\
Instanton  vacuum \cite{nam6} & $1$ & $0.18$ & $0.07$&--\\
\end{tabular}
\end{ruledtabular}
\end{table}
Our predictions for $\langle \xi_{2} \rangle$ and $\langle \xi_{4}
\rangle$ in set I are nearly equal to those of the LFQM and lattice
QCD for $\mu=1~\rm {GeV}$.

To evaluate the differential branching ratio of the $B \to K
\mu^{+}\mu^{-}$ transition on $q^2$, we need to calculate the
transition form factors. The explicit expressions of these form
factors in terms of the light-cone DAs are given in Ref.
\cite{Aliev1}. We use these expressions and replace the holographic
DAs in them; then we convert the obtained form factors based on the
following definitions, which are more conventional \cite{Bailey}:
\begin{eqnarray}\label{eq33}
\langle K(p)|\bar{s}\,\gamma_{\mu}\,b|B(p_{B})\rangle&=&
P_{\mu}\,f_{+}(q^2)+q_{\mu}\,\frac{m_{B}^2-m_{K}^2}{q^2}\,[f_{-}(q^2)-f_{+}(q^2)],\nonumber\\
\langle
K(p)|\bar{s}\,i\,\sigma_{\mu\nu}\,q^{\nu}\,(1+\gamma_{5})\,b|B(p_{B})\rangle&=&
\,[P_{\mu}\,q^{2}-(m_{B}^2-m_{K}^{2})\,q_{\mu}]\,\frac{f_{_T}}{m_{B}+m_{K}}.
\end{eqnarray}
In these definitions, $p$ and $p_{B}$ refer to the momentums of the $K$ and
$B$ meson, respectively; $q=p_{B}-p$ is the momentum carried  by
leptons and $P=p_{B}+q$.

Usually, the numerical results for the form factors calculated via
different methods in QCD have a cutoff. So, to evaluate the form
factors for the whole physical region $m_\ell^2 \le q^2 \le
(m_B-m_{K})^2$,  we look for a good parametrization of the form
factors in such a way that, in the large values of $q^2$, this
parametrization can coincide with the lattice predictions
\cite{Bailey}. Our numerical calculations show that the sufficient
parametrization of the form factors with respect to $q^2$ is as
follows:
\begin{equation}\label{eq34}
F(q^2)=\frac{1}{1- (\frac{q^2}{m_B^2})}\sum_{r=0}^{2} b_r \left[z^r
+ (-1)^{r}\, \frac{r}{3}\, z^4 \right]\,,
\end{equation}
where
$z=\frac{\sqrt{t_{+}-q^2}-\sqrt{t_{+}-t_{0}}}{\sqrt{t_{+}-q^2}+\sqrt{t_{+}-t_{0}}}$,
$t_{+}=(m_{B}+m_{K})^2$, and
$t_{0}=(m_{B}+m_{K})(\sqrt{m_B}-\sqrt{m_{K}})^2$ \cite{Bourrely}.
Table \ref{T36}  shows the values of $b_r~ (r=0,...,2)$ for the form
factors.
\begin{table}[th]
\caption{Results of $z$-expansion fits of the $B\to  K$ form
factors.} \label{T36}
\begin{ruledtabular}
\begin{tabular}{cccccccccccc}
$(A=1;B=1)$&$b_{0}$&  $b_{1}$&$b_{2}$ & $(A=0;B=1)$ &$b_{0}$ &$b_{1}$ &$b_{2}$ \\
\hline
$f_{+}$ &  0.43  &  -1.13  &  -0.21 & $f_{+}$ &  0.38  &  -1.54  & -0.85  \\
$f_{-}$ &   0.27 &  0.08  &   -0.25 & $f_{-}$ &  0.24  &  -0.31  & -1.01  \\
$f_{T}$ &  0.45  & -0.99   &   0.12 & $f_{T}$ &  0.40  & -1.50
&-0.41
\end{tabular}
\end{ruledtabular}
\end{table}

\begin{figure}[!ht]
\includegraphics[width=5 cm,height=6 cm]{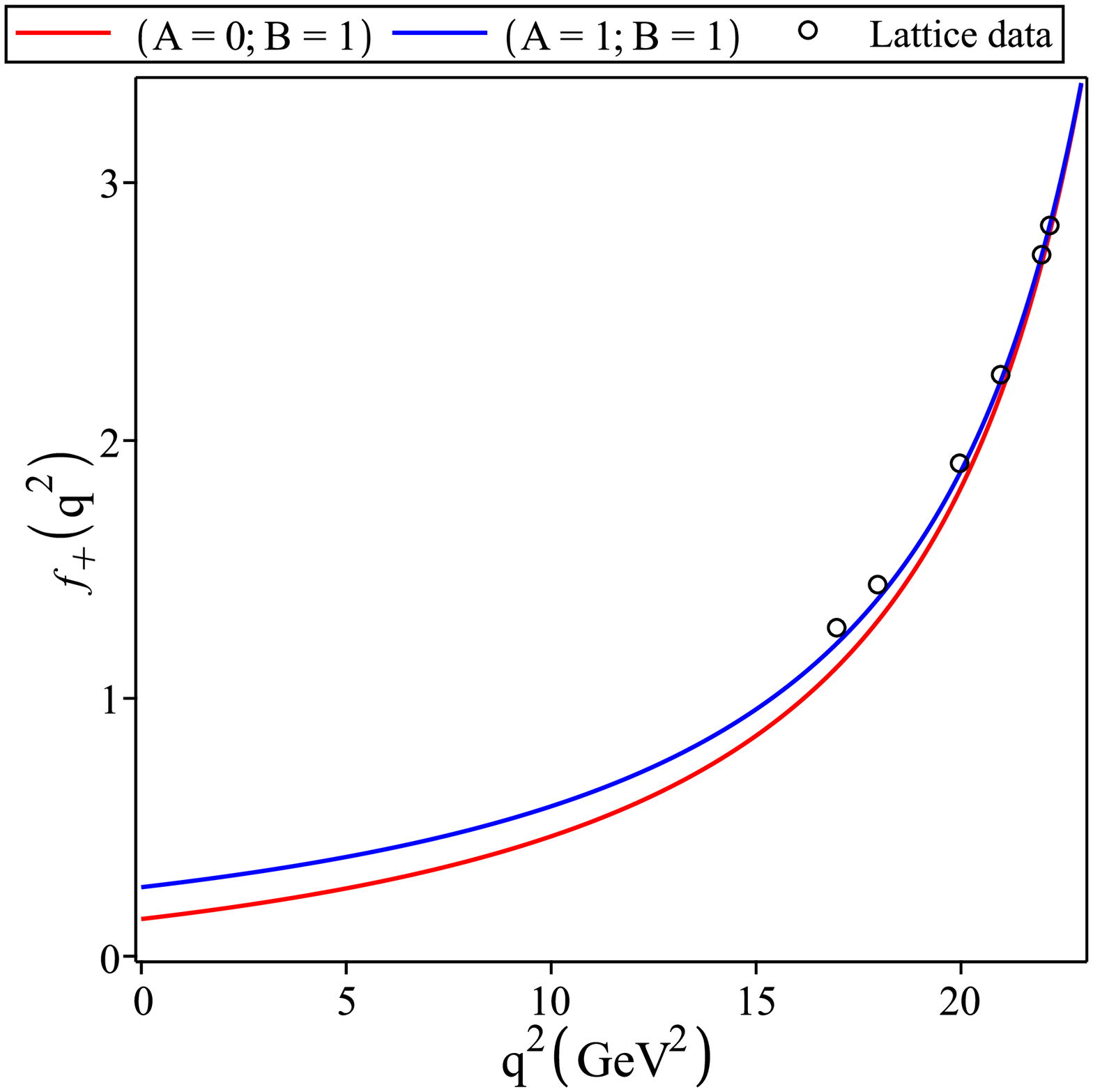}
\includegraphics[width=5 cm,height=6 cm]{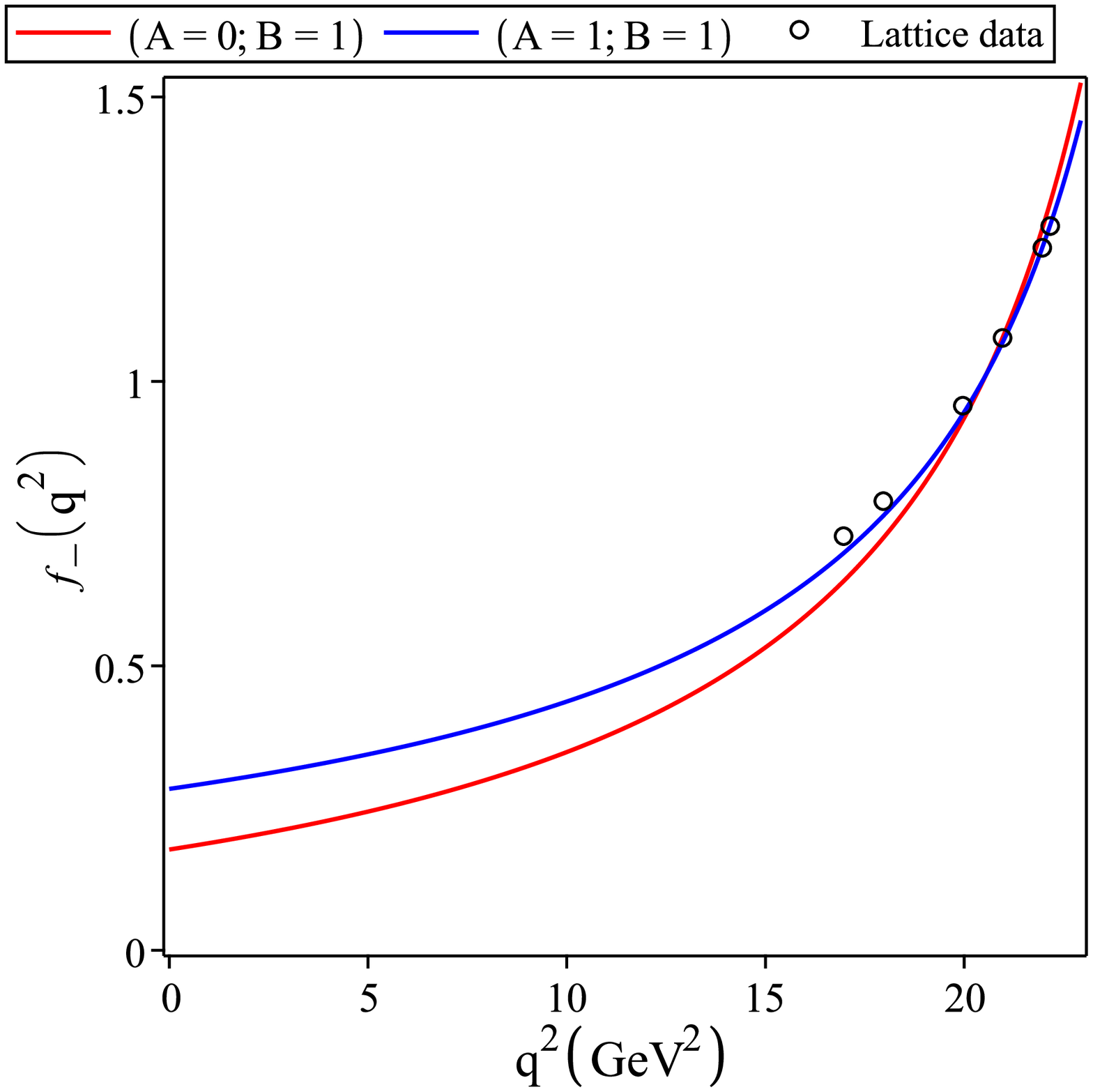}
\includegraphics[width=5 cm,height=6 cm]{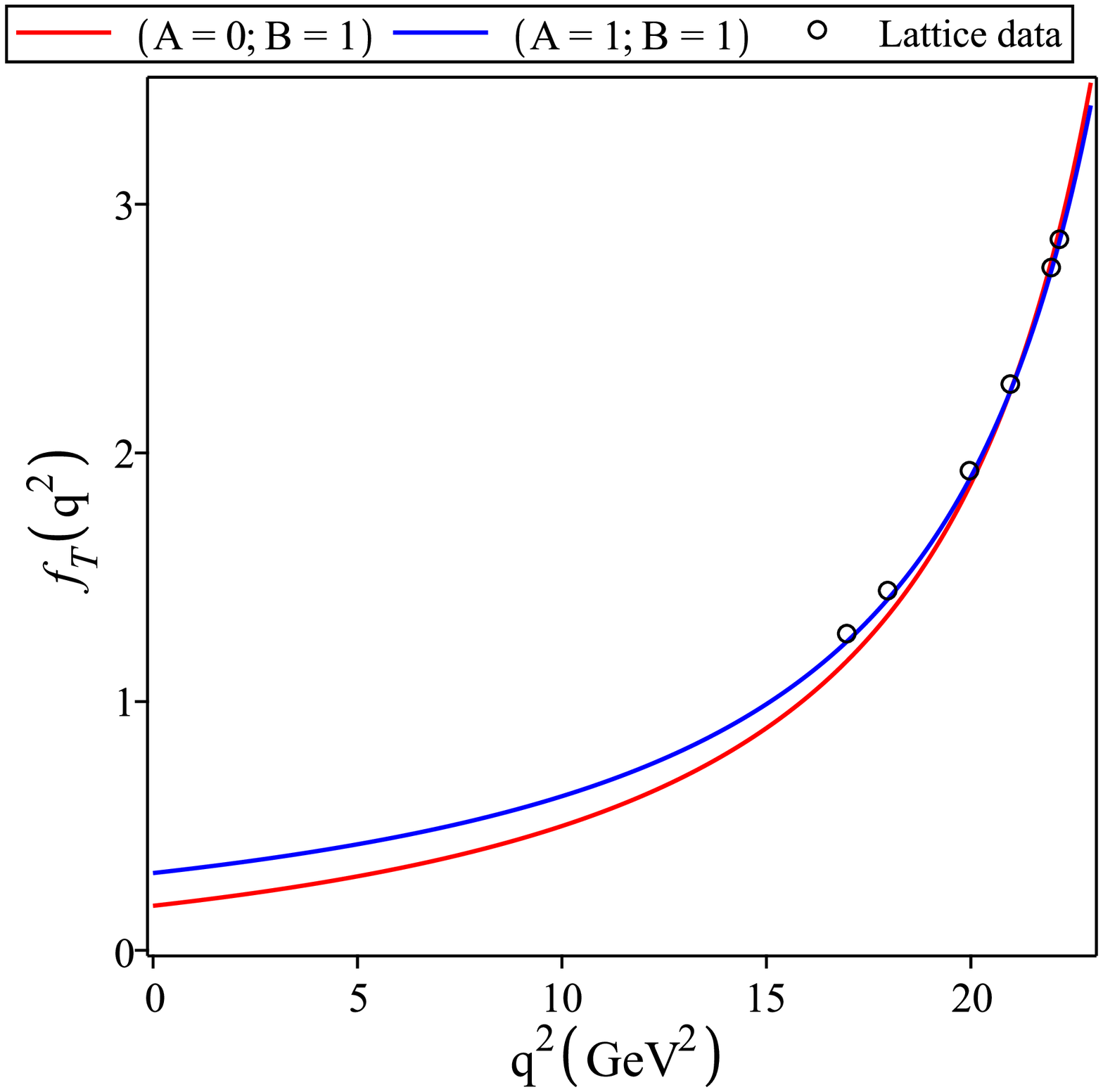}
\caption{The form factor $f_{+}, f_{-}$ and $f_{T}$ of the $B \to K$ decay on $q^2$.
Circles show the lattice data in large $q^2$.}\label{F3}
\end{figure}

Figure \ref{F3} shows the results for the $f_{+}, f_{-}$, and
$f_{T}$ form factor in two sets. In this figure, circles show the
lattice predictions in the large values of $q^2$.

Now, we can evaluate the differential branching ratio of the $B \to
K \mu^{+}\mu^{-}$ transition on $q^2$. The transition of the $B$
meson to the final state $K \mu^+\mu^-$ receives contributions from
tree level decays and decays mediated through virtual quantum loop
processes. The tree level decays proceed through the decay of a $B$
meson to a vector $c\bar{c}$ resonance and a $K$ meson, followed by
the decay of the resonance to a pair of muons. Decays mediated by
FCNC loop processes give rise to pairs of muons with a nonresonant
mass distribution. A broad peaking structure is observed in the
dimuon spectrum of $B \to K \mu^{+}\mu^{-}$ decay in the kinematic
region where the kaon has a low recoil against the dimuon system
\cite{Aaij4}.

In the SM, the semileptonic decays such as the $B \to K \ell^+
\ell^-$ transitions that occur via $b \to s~ \ell^+ \ell^-$
transition, are described by the effective Hamiltonian as
\begin{equation}
H_{\rm eff} = - \frac{G_F}{\sqrt{2}} V_{tb}V_{ts}^{*}
\sum_{i=1}^{10} C_i(\mu)  O_i(\mu)\,,
\end{equation}
where $V_{tb}$ and $V_{ts}$ are the elements of the CKM matrix, and
$C_i(\mu)$ are the Wilson coefficients. The standard set of the
local operators $O_i(\mu)$ is found, for example, in Ref.
\cite{Buras0}. The most relevant contributions to $b \to s~ \ell^+
\ell^-$ transitions are (a) the tree level operators $O_{1,2}$, (b)
the penguin operator $O_{7}$, and (c) the box operators $O_{9,10}$.
The current-current operators $O_{1,2}$ involve an intermediate
charm loop coupled to the lepton pair via the virtual photon (see
Fig. 1). The  electroweak penguin operators $O_{7}$, and $O_{9,10}$
are responsible for the short-distance (SD) effects in the FCNC $b
\to s$ transition, but  the operators $O_{1,2}$  involve both SD and
LD contributions in this transition. In the naive factorization
approximation, contributions of the $O_{1,2}$ operators have the
same form factor dependence as $C_9$ and can, therefore, be absorbed
into an effective Wilson coefficient $C^{\rm eff}_9$ \cite{Lyon}.
\begin{figure}[th]
\includegraphics[width=10cm,height=3cm]{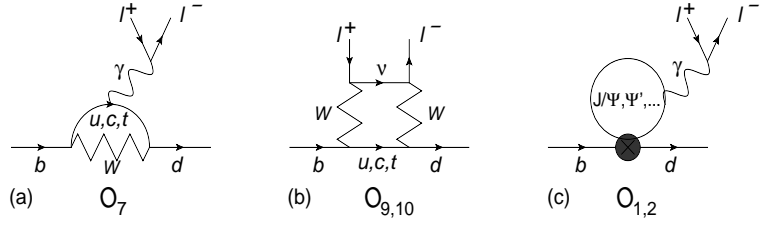}
\caption{(a) and (b) $O_7$ and $O_{9,10}$ short-distance
contributions.   (c) $O_{1,2}$ long-distance charm-loop
contribution.} \label{F21}
\end{figure}
Therefore, the effective Wilson coefficient $C_{9}^{\rm{eff}}$ is
given as  $C^{\rm eff}_9 = C_9 + Y_{SD}(q^2)+Y_{LD}(q^2)$, where
$Y_{SD}(q^2)$ describes the SD contributions from four-quark
operators far away from the resonance regions. The LD contributions,
$Y_{LD}(q^2)$ from four-quark operators near the $c\bar{c}$
resonances cannot be calculated from the first principles of QCD and
are usually parametrized in the form of a phenomenological
Breit-Wigner formula as \cite{Buras0}
\begin{eqnarray}\label{eq026}
Y_{LD}(q^2)&=&\frac{3\pi}{\alpha^2}
\sum_{V_i=\psi(1s),\psi(2s)}\frac{\Gamma(V_i\to l^+
l^-)m_{V_i}}{m_{V_i}^2-q^2-i m_{V_i} \Gamma_{V_i}}.
\end{eqnarray}

\begin{figure}[!ht]
\includegraphics[width=8 cm,height=7 cm]{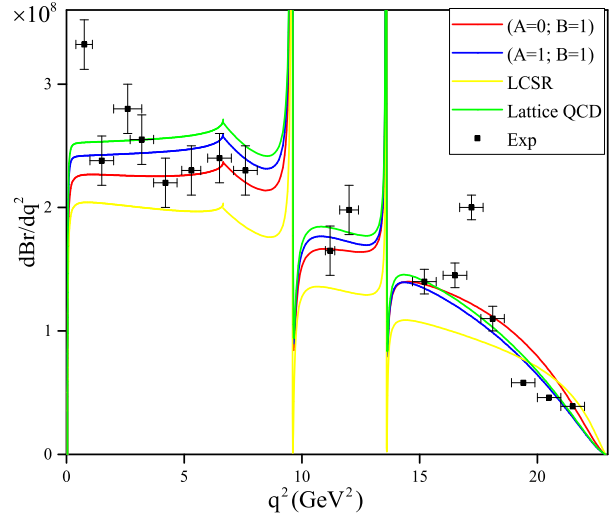}
\caption{The differential branching ratios of the semileptonic $B\to
K\mu^{+}\,\mu^{-}$ decays on $q^2$.}\label{F4}
\end{figure}

The expressions of the differential decay width $d\Gamma/dq^2$ for
the $B \to K\,l^{+}\,l^{-}$ can be found in \cite{Bourrely}. This
expression contains the CKM matrix elements, Wilson coefficients,
and form factors related to the definitions in Eq. (\ref{eq33}). In
this paper, we take $C_{ 7}^{\rm eff}=-0.313$, $C_{10}=-4.669$
\cite{Faessler} and use $C_{ 9}^{\rm eff}$ according to Ref.
\cite{Buras0}. Considering two charm resonances, $\psi(1s)$ and
$\psi(2s)$, the dependency of the differential branching ratio for
the $B \to K \mu^{+}\mu^{-}$ decay on $q^2$ is presented in Fig.
\ref{F4}. In this figure, the results obtained by the LCSR
\cite{Aliev1} and lattice QCD \cite{Bailey} approaches are shown
with yellow and green lines, respectively. Also, the experimental
values \cite{Bifani} with their errors are plotted in this figure.
As can be seen in Fig. \ref{F4}, the predictions of all models for
the differential branching ratio of the $B \to K \mu^{+}\mu^{-}$
transition on $q^2$ are not in a good agreement with the
experimental value in the low energy region ($q^2 < 1~ \rm GeV^2$)
where the nonperturbative QCD overcomes. For the momentum transfer
squared between $1~ \rm GeV^2 < q^2 < 10~ \rm GeV^2$, a large number
of the experimental values (central values) are between our
predictions via the AdS/QCD correspondence for two sets. In the high
momentum transfer squared region $(q^2 > 10 ~ \rm GeV^2)$, the
predictions of the lattice QCD and AdS/QCD for two sets, are well
fitted to experimental values (by considering their errors).

To  summarize, based on the dynamical spin effects, we extracted the
twist-$2$, $-3$, and $-4$ DAs of the $K$ pseudoscalar meson in the
AdS/QCD correspondence approach. The AdS/QCD scale $\kappa = 590~
\rm{MeV}$;  this value is provide by fitting it to the Regge slopes,
and two sets $(A=1;B=1)$ and $(A=0;B=1)$  for the dynamical spin
effects were used in our analysis. For a better analysis, we
calculated the masses of the light quarks with the help of the
experimental values for  the decay constants of pion and kaon
pseudoscalar mesons in two sets. The radius, and the EM form factor
of the $K$ meson, quantities related to the holographic LFWF
$\Psi^K(r, \alpha)$, were investigated and compared  with the
lattice QCD and experimental values. By evaluating the transition
form factors with the help of the holographic DAs, the differential
branching ratio for the $B \to K \mu^+ \mu^- $ decay on $q^2$ was
plotted for two sets of $A$ and $B$. A comparison with the
experimental values showed that our predictions with the AdS/QCD
correspondence were good.

\section*{ACKNOWLEDGMENTS}
Partial support from the Isfahan University of Technology Research
Council is appreciated.

\end{document}